\documentclass[twoside,fleqn]{article}
\usepackage{epsfig}
\usepackage{espcrc2}

\title{ Thermodynamics of two-colour QCD \thanks{ This work was partly
    supported by the Deutsche Forschungsgemeinschaft under grant Ka 1198/3-1
    and the EU TMR network grant ERBFMRX-CT97-0122.} }

\author{ O. Kaczmarek with F. Karsch and E. Laermann \medskip\\
Fakult\"at f\"ur Physik, Universit\"at Bielefeld, 33501 Bielefeld, Germany }

\begin{document}

\begin{abstract}
We discuss the thermodynamics of two-colour QCD with four flavours
of staggered quarks on $8^3\times4$ and $16^3\times4$ lattices.
In our simulations we use the Naik action for the fermions and a
(1,2) tree-level improved
gauge action. We analyze the deconfinement
and chiral phase transitions for four different quark masses
(m=0.1,0.05,0.025,0.015). Contrary to three-colour QCD the peak in the
Polyakov loop susceptibility decreases with decreasing quark
mass. This reflects an early breaking of the string in the
heavy quark potential, which we verify explicitly
by calculating the heavy quark potential at finite temperature
using Polyakov loop correlations.
\end{abstract}

\maketitle

\section{Introduction}
\vspace{-0.8ex}
          We investigate the thermodynamics of QCD with four flavours and
          colour group $SU(2)$
          in the staggered discretisation \cite{heller}. In the continuum limit
          the phase transition
          is expected to be of first order induced by fluctuations
          \cite{pisarski}.

          In our simulations we used the exact hybrid-MC algorithm with a
          Symanzik improved $(1,2)$ gluon action and the
          Naik improvement for
          the fermionic sector. To reduce the number of conjugate gradient
          steps we
          used the minimal residual extrapolation method
          (MRE)\cite{brower}. The residuum for the CG was $10^{-8}$ inside the molecular
          dynamic trajectory and $10^{-13}$ elsewhere. Each trajectory was of length
          1, while the number of steps in the molecular dynamic evolution was adjusted to
          get a good acceptance.

          For the analysis of the phase transition lattices of size $8^3\times4$
          with three different bare quark masses $m=0.1, 0.05, 0.025$ and
          size $16^3\times4$ with $m=0.015$ were
          used to calculate the Polyakov loop, the chiral condensate and their
          susceptibilities. We used a noisy estimator with 25 random sources
          to calculate the susceptibility of $\bar\psi\psi$.
          Simulations on $16^3\times4$ lattices at $m=0.025$ were
          performed to calculate finite temperature potentials from
          Polyakov loop correlations.
          For comparison of these potentials we also used results of two flavour QCD with
          colour group $SU(3)$ and standard gluonic and fermionic actions \cite{edwin}.

\section{Polyakov Loop}
\vspace{-0.8ex}
         The results for the Polyakov loop are shown in Fig.1. With decreasing
          mass they seem to get flatter showing no clear sign of a
          phase-transition. For small $\beta$-values the Polyakov loop does not
          drop to zero which is a first hint for string breaking.
          \begin{figure}[htbp]
            \label{pol}
            \epsfig{file=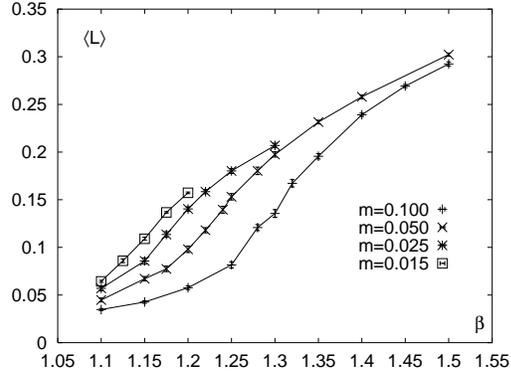, width =7.2cm}
            \vspace*{-5ex}
            \caption[]{Polyakov loop on $8^3\times4$ lattices (m=0.1, 0.05,
              0.025) and on $16^3\times4$ lattices (m=0.015).}
          \end{figure}
          Comparing the data for the susceptibility of the Polyakov loop for
          $SU(2)$ with four flavours (Fig.2) and $SU(3)$ with two flavours \cite{karsch}
          one sees quite a different mass dependence of the peak heights. For $SU(2)$
          the peaks decrease with decreasing mass while for $SU(3)$ the heights
          have the tendency to increase slightly. This reflects a smooth crossover
          from the low to the high temperature phase and an early breaking of
          the string for $SU(2)$ (see Fig.5).
          \begin{figure}[htbp]
            \label{sus}
            \epsfig{file=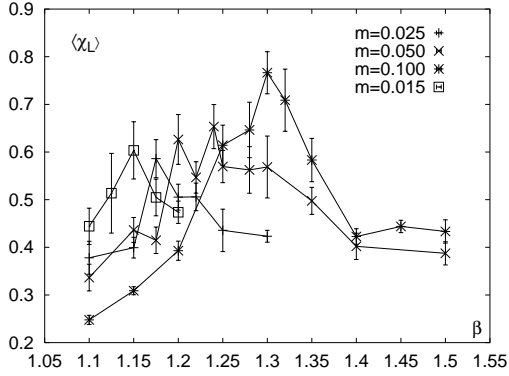, width =7.2cm}
            \vspace*{-5ex}
            \caption[]{Susceptibility of the Polyakov loop on $8^3\times4$ lattices (m=0.1, 0.05,
              0.025) and on $16^3\times4$ lattices (m=0.015).}
          \end{figure}
          Further simulations are needed to see if this behaviour is also valid
          for larger $N_\tau$ or if it could be a strong coupling effect.
\vspace*{-1ex}
\section{Chiral Condensate}
\vspace*{-1ex}
          \begin{figure}[htbp]
            \label{chi}
            \epsfig{file=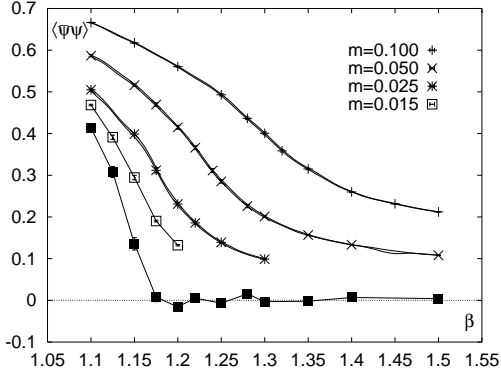, width =7.2cm}
            \vspace*{-5ex}
           \caption[]{Chiral condensate for m=0.1, 0.05, 0.025 ($8^3\times4$),
             m=0.015 ($16^3\times4$) and the linear extrapolation to m=0.
               }
          \end{figure}
          The data for the chiral condensate in Fig.3 show a clear
          phase transition in the chiral limit
          although $\langle\bar\psi\psi\rangle$ stays finite for finite bare quark
          masses.
          The results for the two smallest masses were used
          to extrapolate to the zero mass limit with a linear ansatz
          \begin{eqnarray}
            \langle\bar\psi\psi\rangle(m=0) = \langle\bar\psi\psi\rangle(m) - c\cdot m.
          \end{eqnarray}
          For $\beta>1.2$ chiral symmetry is restored in
          the zero mass limit within errors. A wider range of masses are needed
          to use other ans\"atze for the extrapolation.

          So far, our results are compatible with earlier investigations by
          Kogut et al. \cite{wyld}.
          \begin{figure}[htbp]
            \label{chi2}
            \epsfig{file=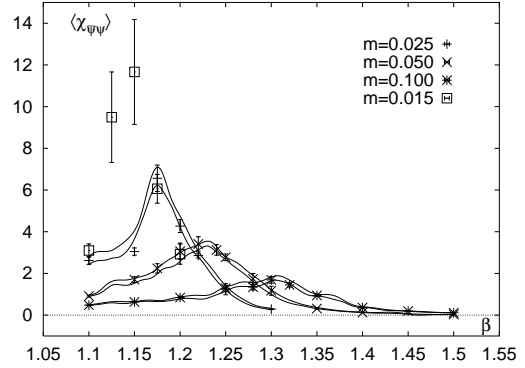, width =7.2cm}
            \vspace*{-5ex}
            \caption[]{Susceptibility of the chiral condensate for m=0.1, 0.05, 0.025 ($8^3\times4$),
             m=0.015 ($16^3\times4$).
               }
          \end{figure}
          In contrast to the susceptibility of $L$ the comparison of $SU(2)$
          and $SU(3)$ data \cite{karsch} shows no major difference for the susceptibility of
          the chiral condensate. Both exhibit
          clear peaks that increase with decreasing bare
          quark mass. Also quantitatively the quark mass dependence of the peak
          heights is quite similar for both gauge groups. We have fitted the
          peak heights to the ansatz
          \begin{eqnarray}
            \chi_{\bar\psi \psi,max}(m) \sim m^{-z_m}.
          \end{eqnarray}
          \begin{table}[htbp]
            \begin{center}
              \begin{tabular}{|c|c|c|}
                \hline
                Action & Lattice & $z_m$\\
                \hline
                $SU(2),N_f=4$ & $8^3\times 4$ & 0.95(7)\\
                $SU(3),N_f=2$ & $8^3\times 4$ & 0.84(5)\\
                $SU(3),N_f=2$ & $16^3\times 4$ & 0.93(8)\\
                \hline
              \end{tabular}
            \end{center}
            Table 1. Fit-results of ansatz (2).
        \end{table}

          The results of the fits for $SU(2)$ and $SU(3)$ are shown in Tab.1.
          Although we do not expect a second order transition for 4-flavour,
          $N_c=2$ QCD in the continuum limit, we find no indication for a
          discontinuity, i.e. a first order phase transition. This may indicate
          that the transition is still controlled by the
          $U(1)$ symmetry of the staggered fermion action, i.e. we are still at
          too strong coupling.
\vspace*{-1.5ex}
\section{Finite Temperature Potentials}
\vspace*{-0.5ex}
          We have calculated finite temperature potentials on $16^3\times4$
          lattices for $m=0.025$ and $T<T_c$
          using Polyakov loop correlations up to a distance of $RT=2.0$
          \begin{eqnarray}
            PLC(R)=\langle L^\dagger(0,0,0)
            L(dx_1,dx_2,dx_3) \rangle
          \end{eqnarray}
           with $dx_i=0..4$ and on-axis
            $dx_i=0..8$.

          The potentials are extracted from the following relation:
          \begin{eqnarray}
            V(R)=-\frac{1}{N_\tau} \log \langle PLC(R) \rangle.
          \end{eqnarray}
          \begin{figure}[htbp]
            \label{plc}
            \epsfig{file=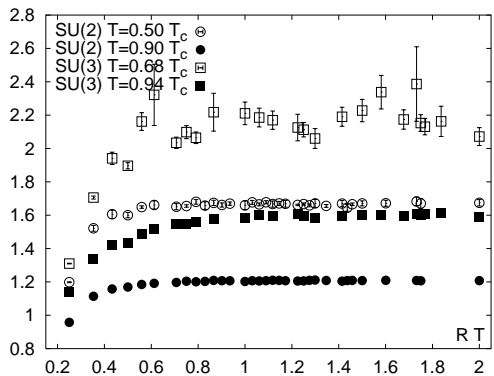, width =7.2cm}
            \vspace*{-5ex}
            \caption[]{Comparison of the Potentials $V(R)$ for $N_f=4-SU(2)$ and
                $N_f=2-SU(3)$ for $T\approx T_c$ and smaller temperatures.
              }
          \end{figure}

          We compare our potentials with results for two flavour QCD \cite{edwin}.
          For both theories we see clear
          signs of string breaking. The potentials seem to rise
          up to a critical distance $R_c$ where the string breaks and
          $V(R>R_c)$ stays constant. It seems that for $SU(2)$
          the string breaks
          at smaller distances than for $SU(3)$ which may be attributed
          to the larger number of light
          states in the $SU(2)$-theory that can contribute to the string
          breaking.
          It is however not yet clear how the heights of
          the potentials are controlled by self-energy contributions.

\section{Conclusions}
          So far we have no indications for a first order phase transition for
          QCD with four flavours and colour group $SU(2)$. This may be a strong
          coupling artifact due to the incomplete chiral symmetry of the
          staggered action.
          Further investigations with larger temporal
          extent are needed to get closer to the continuum limit.

          A rather striking feature of the $SU(2)$ theory is the decoupling of
          chiral symmetry restoring and deconfining aspects of the
          transition. While the chiral order parameter changes more rapidly
          close to $T_c$ with decreasing quark mass the Polyakov loop
          expectation values become flatter in this region.

          Our data are compatible with earlier results by Kogut et
          al. \cite{wyld}. Simulations at smaller quark masses should be performed to
          see if a first order phase transition appears as suggested by Kogut \cite{kogut}.

          The finite temperature potentials show signs of string breaking for
          $SU(2)$ in agreement with the $SU(3)$ data of C. DeTar et al. \cite{edwin}.
          Due to the larger number of light states in $SU(2)$ it seems
          that the string breaks earlier and the plateaus are lower than for
          $SU(3)$.

          So far this is a qualitative comparison.
          One has to analyze the self-energy contributions to the potentials
          and needs information about the lattice spacing to really compare
          both.

\end{document}